\documentclass[english]{article}
\usepackage[T1]{fontenc}
\usepackage[latin9]{inputenc}
\usepackage{xcolor}
\usepackage{float}
\usepackage{amssymb}
\usepackage{stackrel}
\usepackage{graphicx}
\usepackage{esint}

\makeatletter

\pdfoutput=1
\usepackage[T1]{fontenc}
\usepackage[latin9]{inputenc}
\usepackage{color}
\usepackage{float}
\usepackage{graphicx}
\usepackage{esint}
\usepackage{enumerate}
\usepackage{tikz}

\makeatletter

\usepackage{amsfonts,amssymb}
\usepackage{latexsym}
\usepackage{epsfig}
\usepackage{cite}
\usepackage{graphicx}
\usepackage[colorlinks,linkcolor=blue]{hyperref}
\newcommand{\be}{\begin{equation}}
\newcommand{\ee}{\end{equation}}

\usepackage{babel}

\makeatother

\usepackage{babel}
\begin{document}
{}~ \hfill\vbox{\hbox{ }}\break
\vskip 3.0cm
\centerline{\Large \bf Three dimensional regular black string via loop corrections}

\vspace*{10.0ex}
\centerline{\large Shuxuan Ying}
\vspace*{7.0ex}
\vspace*{4.0ex}
\centerline{\large \it Department of Physics, Chongqing University}
\vspace*{1.0ex}
\centerline{\large \it Chongqing, 401331, China} \vspace*{1.0ex}
\vspace*{4.0ex}

\centerline{ysxuan@cqu.edu.cn}
\vspace*{10.0ex}
\centerline{\bf Abstract} \bigskip \smallskip

It is well known that some spacetime singularities of low energy effective action can be resolved by the effective loop corrections. However, the known potentials for the loop corrections fail to regulate the singularity of three dimensional black string. An alternative method is to introduce an extra flat spatial direction and boost away the singularity by using the $O\left(d,d\right)$ rotation with the special Kalb-Ramond field.  In this paper, we investigate a new set of non-local dilaton potentials for the effective loop corrections. The result shows that the singularity of three dimensional black string can be consistently resolved by the loop corrections.

\vfill
\eject
\baselineskip=16pt
\vspace*{10.0ex}
\tableofcontents

\section{Introduction}

T-duality in string theory has attracted a lot of attention in recent years. It indicates the physical equivalence between non-contractible
closed strings winding on two compactified target
space with different radii. For
the uncompactified FLRW background $g_{\mu\nu}=\mathrm{diag}\left(-1,a\left(t\right)^{2}\delta_{ij}\right)$,
T-duality becomes the continuous $O\left(d,d\right)$ symmetry which manifests the scale factor duality $a\left(t\right)\rightarrow1/a\left(t\right)$ in the low energy effective action
 \cite{Tseytlin:1991wr,Veneziano:1991ek,Meissner:1991zj, Sen:1991zi,Sen:1991cn,Tseytlin:1991xk}.
This property does not exist in Einstein's gravity, since the closed
string dilaton $\phi$ plays a central role in this duality. Moreover, the low energy effective action only applies to the
perturbative region.
Beyond the perturbative region, there are two types of corrections
which need to be included in the action. One is the higher order curvature
corrections, which is based on the ratio  between the string length $\sqrt{\alpha^{\prime}}$
and the scale curvature of the background. When scale curvature
is small enough, the string effect dominates. Therefore,
the $\alpha^{\prime}$ corrections need to be considered in the action. The second kind of corrections comes from the higher order genus expansions. It
is derived from the change of topological complexity of string
worldsheet. The complexity is determined by the string coupling constant: $g_{s}=\exp\left(\phi\right)$. These corrections are known as
loop corrections.

In recent progress of $\alpha^{\prime}$ corrections, Hohm and Zwiebach
find that all order $\alpha^{\prime}$ corrections can be dramatically
simplified as a summation of the Hubble parameters in the FLRW spacetime
by using the $O\left(d,d\right)$\textcolor{brown}{{} }duality \cite{Hohm:2019ccp,Hohm:2019jgu}.
It is then worth to study the exact solutions nonperturbatively to all order
$\alpha^{\prime}$ corrections. Some possible regular solutions have
been found in refs. \cite{Wang:2019mwi,Wang:2019kez,Ying:2021xse,Ying:2022xaj}.
The discussions about the matter sources can be found in ref. \cite{Bernardo:2019bkz}.
On the other hand, there is no significant progress of loop
corrections.
The reason is that it is difficult to determine the coefficients of loop corrections.
To see the known loop corrected formalism, let us recall the
action \cite{Gasperini:2003pb,Gasperini:2004ss}
\begin{equation}
S=-\frac{1}{2{\lambda_{s}^{D-2}}}\int d^{D}x\sqrt{-g}e^{-2\phi}\left(R+4\left(\partial_{\mu}\phi\right)^{2}-2{\lambda_{s}^{D-2}}V\left(e^{-\Phi}\right)\right),\label{eq:action with potential}
\end{equation}

\noindent where $\phi$ is a physical dilaton, $\Phi$ denotes an
 $O\left(d,d\right)$ invariant non-local dilaton, $V\left(e^{-\Phi}\right)$ is an effective dilaton potential which is used to mimic the backreaction of the loop corrections from the compactified higher dimensional manifold,   $\lambda_{s}^{2}=2\pi\alpha^{\prime}$ and the potential
$V$ possesses the canonical dimensions of an energy density $\left[V\right]=\left(mass\right)^{D}$.
In \cite{Gasperini:2003pb,Gasperini:2004ss},
the authors made numerous attempts to guess a general non-local
dilaton potential:
\begin{equation}
V\left(e^{-\Phi\left(x\right)}\right)=m^{2}e^{2\Phi\left(x\right)}\left[\left(\beta-e^{2n\Phi\left(x\right)}\right)^{\frac{2n-1}{n}}-d\right],\label{eq:previous loop potential}
\end{equation}
where $\beta$ is a dimensionless coefficient, $n$ is a
\textquotedblleft loop-counting\textquotedblright{} parameter, $d$ is the number of spatial dimension, the parameter $m$ has the dimension of $\left(mass\right)^{D/2}$ and
the non-local dilaton $\Phi\left(x\right)$ is defined by

\begin{equation}
e^{-\Phi\left(x\right)}={\lambda_{s}^{-\left(D-1\right)}}\int d^{D}x^{\prime}\sqrt{-g\left(x^{\prime}\right)}e^{-2\phi\left(x^{\prime}\right)}\sqrt{-4g^{\mu\nu}\partial_{\mu}\phi\left(x^{\prime}\right)\partial_{\nu}\phi\left(x^{\prime}\right)}\delta\left(2\phi\left(x^{\prime}\right)-2\phi\left(x\right)\right),
\end{equation}

\noindent The ``non-local'' means that this function is defined
by the non-local function of the physical dilaton $\phi$. This definition is necessary
 since the action must be invariant under the $O\left(d,d\right)$
transformations and diffeomorphism simultaneously. Moreover,
in the cosmology or black string background,
it reduces to $e^{-\Phi\left(x\right)}={\lambda_{s}^{-\left(D-1\right)}}V_{D-1}\sqrt{g}e^{-2\phi}$
which is dimensionless as expected.
We also call $\Phi\left(x\right)$
as an $O\left(d,d\right)$ dilaton in the rest of the paper. Based on
this non-local dilaton potential (\ref{eq:previous loop potential}), the big-bang singularity can be
resolved in the non-perturbative regions. After
removing the singularity, the pre- and post-big bang evolutions
are smoothly connected. Then, the universe evolves from the string
perturbation vacuum $g_{s}\rightarrow0$ to the strong string coupling
region $g_{s}\sim1$. The other regular solutions of cosmology and black holes can be found in \cite{Gasperini:1992em,Veneziano:2000pz,Gasperini:2002bn,Gasperini:2003pb,Gasperini:2004ss,Gasperini:2007vw}.
However, there is an exception here; The non-local dilaton potential fails to resolve the singularity
of three dimensional black string \cite{Horne:1991gn} or two dimensional
Witten's black hole \cite{Mandal:1991tz,Lemos:1993py,Witten:1991}. The alternative
method is to add an extra flat spatial direction and boost away the singularity by introducing the Kalb-Ramond
field through the $O\left(d,d\right)$ rotation \cite{Gasperini:1992ym}.
Inspired by a map between
$\alpha^{\prime}$ corrections and loop corrections \cite{Wang:2019dcj}, we construct the most general dilaton
potential for the loop corrections in this paper:

\begin{equation}
2{\lambda_{s}^{D-2}}V\left(e^{-\Phi\left(x\right)}\right)=-\lambda^{2}+\stackrel[k=1]{\infty}{\sum}b_{k}\left(\gamma\right)e^{2k\Phi\left(x\right)},
\end{equation}

\noindent where $b_{k}\left(\gamma\right)$'s are undetermined constants, $\gamma$ is a coupling constant and  $\lambda^{2}=-\frac{2\left(D-26\right)}{3\alpha^{\prime}}$ which corresponds to the cosmological constant and covers the tree-level low energy effective action.
In this construction, $e^{-\Phi\left(x\right)}$ is also defined by the non-local dilaton.
When $\gamma\rightarrow0$,
$b_{k}\left(0\right)$ becomes zero, and then the loop effects disappear.
Adding this potential into the low energy effective action, we find
the regular solution of three dimensional black string. This new solution
does not possess the spacetime singularity but has an event horizon.
Moreover, in the perturbative region, this solution covers the perturbative
solution of three dimensional black string. Using this new potential,
it is possible to get more regular solutions in the near future. In this paper, we will set $2\lambda_{s}^{D-2}=1$ for simplicity.

The reminder of this paper is outlined as follows. In section 2, we
briefly review the three dimensional black string solution. In section
3, we construct the most general non-local dilaton potential and find
the regular solution of black string. Section 4 is discussion.

\section{A brief review of three dimensional black string}
In this section, we present a brief review of three dimensional black string. At first, let us recall the three dimensional low energy effective action of closed
string with a cosmological constant:

\begin{equation}
S=\int d^{3}x\sqrt{-g}e^{-2\phi}\left(R+4\left(\nabla\phi\right)^{2}+\lambda^{2}\right),\label{eq:original action}
\end{equation}

\noindent where $g_{\mu\nu}$ is the string metric, $\phi$ denotes the
physical dilaton, the cosmological constant $\lambda^{2}=-\frac{2\left(D-26\right)}{3\alpha^{\prime}}$
and we set the Kalb-Ramond field $b_{\mu\nu}=0$ for
simplicity. The black string solution of this action was first discovered
by Horne and Horowitz \cite{Horne:1991gn}, which is given by

\begin{eqnarray}
ds^{2} & = & -\left(1-\frac{M}{r}\right)dt^{2}+\left(1-\frac{M}{r}\right)^{-1}\frac{1}{\lambda^{2}r^{2}}dr^{2}+d\varphi^{2},\nonumber \\
\phi & = & -\frac{1}{2}\ln\left(\frac{2}{M}r\right),\label{eq:3D black string metric}
\end{eqnarray}

\noindent where $\varphi$ is the compact coordinate. Moreover, it is called black
string since it possesses the asymptotically flat spacetime
and its topology is $R^{2}\times S^{1}$. If we remove the direction
$d\varphi^{2}$ in this metric, it becomes the two dimensional black
hole \cite{Mandal:1991tz}, and it is also known as
Witten's 2D black hole solution which can be obtained from the $SL\left(2,R\right)/U\left(1\right)$
gauged WZW model \cite{Witten:1991}.
Although we only study the black
string solution in the rest of the paper, our results are applicable to both cases and
share the same results. In the metric (\ref{eq:3D black string metric}),
the event horizon locates at $r=M$. Due to the scalar
curvature $R_{0}=\frac{\lambda^{2}M}{r}$, the curvature singularity locates
at $r=0$. On the other hand, there are two kinds
of coordinate transformations which cover the different regions of
maximally extended spacetime. The first one is

\begin{equation}
\frac{r}{M}=\cosh^{2}\left(\frac{\lambda}{2}x\right),\label{eq:coordinates transformation}
\end{equation}

\noindent where $r\geq M$ and $x\geq0$. Utilize this coordinate
transformation, the metric (\ref{eq:3D black string metric}) becomes

\begin{eqnarray}
ds_{I}^{2} & = & g_{00}dt^{2}+g_{11}dx^{2}+g_{22}d\varphi^{2},\nonumber \\
 & = & -\tanh^{2}\left(\frac{\lambda}{2}x\right)dt^{2}+dx^{2}+d\varphi^{2},\nonumber \\
\Phi & = & -\ln\left(\sinh\left(\lambda x\right)\right).\label{eq:O(2,2) metric}
\end{eqnarray}

\noindent Since the metric $g_{\mu\nu}$ depends on the $x$, the
$O\left(d,d\right)$ invariant dilaton is defined by the components
of the metric except the component $g_{11}$ of the direction $dx$ \cite{Wang:2019mwi}:

\begin{equation}
\Phi=2\phi-\ln\sqrt{\left|g_{00}\times g_{22}\right|}.\label{eq:odd invariant dilaton}
\end{equation}

\begin{figure}[H]
\begin{centering}
\includegraphics[scale=0.5]{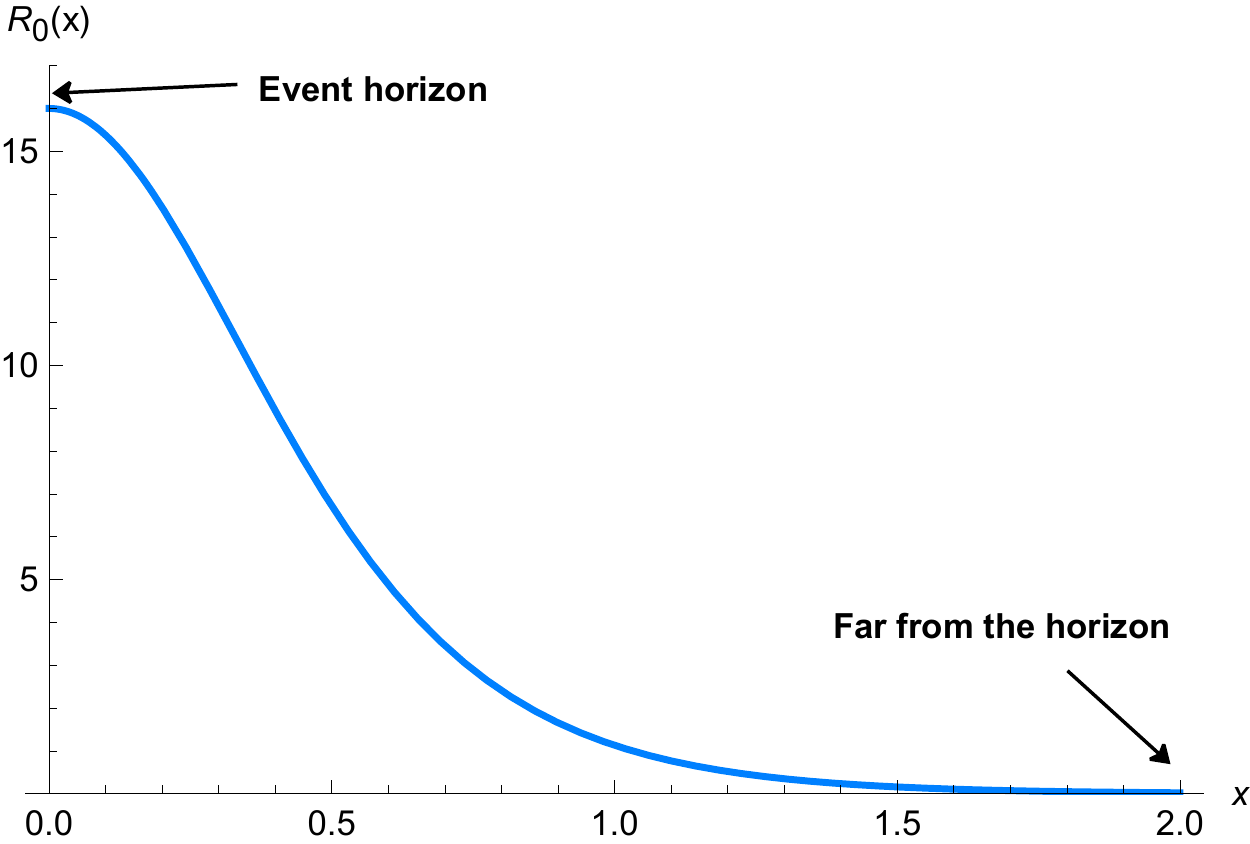}
\par\end{centering}
\centering{}\caption{\label{fig:scalar curvature witten1} The behavior of the scalar curvature
$R_{0}$, where $\lambda=4$ and $x\protect\geq0$.}
\end{figure}

\noindent  In this new metric (\ref{eq:O(2,2) metric}),
it does not possess a curvature singularity since it only describes
the region on the outside of the event horizon ($x=0$), and the scalar
curvature $R_{0}=\lambda^{2}\cosh^{-2}\left(\frac{\lambda x}{2}\right)$
 is regular everywhere in this region, see figure (\ref{fig:scalar curvature witten1}).

\begin{figure}[H]
\begin{centering}
\includegraphics[scale=0.5]{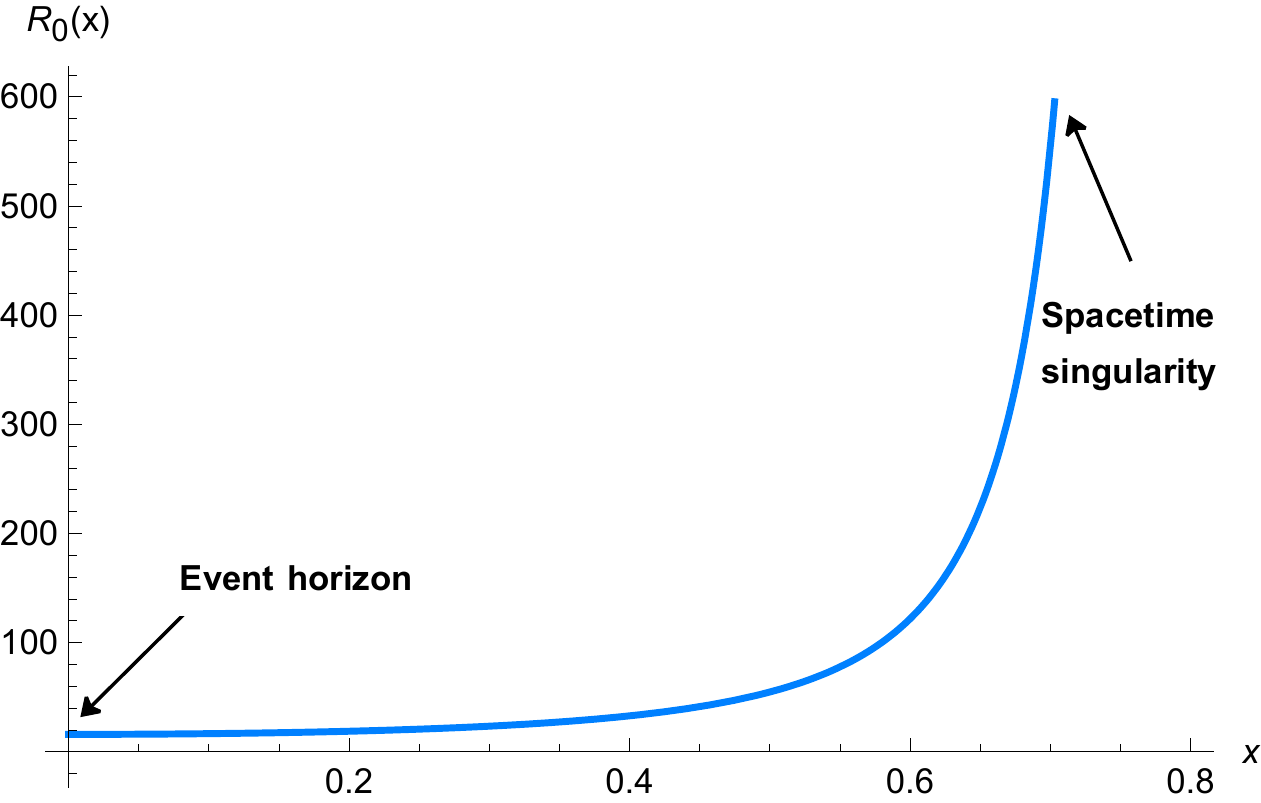}
\par\end{centering}
\centering{}\caption{\label{fig:scalar curvature witten2} The behavior of the scalar curvature
$R_{0}$, where $\lambda=4$ and $0\protect\leq x\protect\leq\frac{\pi}{\lambda}\simeq0.8$.}
\end{figure}

\noindent To study the curvature singularity of the metric (\ref{eq:3D black string metric}),
we need to adopt the second kind of coordinate transformation:

\begin{equation}
\frac{r}{M}=\cos^{2}\left(\frac{\lambda}{2}x\right),
\end{equation}

\noindent where $0\leq r\leq M$ and we only consider one period,
namely $0\leq x\leq\frac{\pi}{\lambda}$. Based on this transformation,
the metric (\ref{eq:3D black string metric}) becomes

\begin{eqnarray}
ds_{II}^{2} & = & g_{11}dx^{2}+g_{00}dt^{2}+g_{22}d\varphi^{2},\nonumber \\
 & = & -dx^{2}+\tan^{2}\left(\frac{\lambda}{2}x\right)dt^{2}+d\varphi^{2},\nonumber \\
\Phi & = & -\ln\left(\sin\left(\lambda x\right)\right),\label{eq:inner metric}
\end{eqnarray}

\noindent which describes the inner metric of black hole, and $x$
here denotes the time-like direction. This metric (\ref{eq:inner metric})
topologically corresponds to a disk. Moreover, the event horizon locates at
$x=0$ and the curvature singularity locates at the boundary of disk
$x=\frac{\pi}{\lambda}$ due to the scalar curvature $R_{0}=\lambda^{2}\cos^{-2}\left(\frac{\lambda x}{2}\right)$,
see figure (\ref{fig:scalar curvature witten2}). Therefore, in this paper, our aim
is to remove the curvature singularity of the metric (\ref{eq:inner metric})
by constructing the non-local dilaton potential.

\section{Regular black string solution by loop corrections}

In this section, we
plan to remove the curvature singularity of the metric (\ref{eq:inner metric})
by loop corrections. To achieve this aim, let us consider the low
energy effective action with the non-local dilaton potential. The
action is given by

\begin{equation}
S=\int d^{D}x\sqrt{-g}e^{-2\phi}\left[R+4\left(\partial_{\mu}\phi\right)^{2}-V\left(e^{-\Phi}\right)\right],\label{eq:action with potential}
\end{equation}

\noindent where $\phi$ is the physical dilaton and $\Phi$ denotes the
 $O\left(d,d\right)$ dilaton. Then, we construct
the most general formula for the non-local dilaton potential, which
includes the arbitrary higher-order loop corrections
\begin{equation}
V\left(e^{-\Phi}\right)=-\lambda^{2}+\stackrel[k=1]{\infty}{\sum}b_{k}\left(\gamma\right)e^{2k\Phi},\label{eq:loop potential}
\end{equation}
where the non-local function $e^{-\Phi}$ \cite{Gasperini:1992em,Gasperini:2003pb}
is given by

\begin{equation}
e^{-\Phi\left(x\right)}=\int d^{D}x^{\prime}\sqrt{-g\left(x^{\prime}\right)}e^{-2\phi\left(x^{\prime}\right)}\sqrt{-4g^{\mu\nu}\partial_{\mu}\phi\left(x^{\prime}\right)\partial_{\nu}\phi\left(x^{\prime}\right)}\delta\left(2\phi\left(x^{\prime}\right)-2\phi\left(x\right)\right).
\end{equation}

\noindent This function is a scalar under the general
coordinate transformations and the $O\left(d,d\right)$ transformations.
Considering the homogeneous and isotropic background, the non-local
function reduces to the following formula:

\begin{equation}
e^{-\Phi\left(x\right)}=V_{D-1}\int dx_{a}^{\prime}\frac{d\left(2\phi\right)}{dx_{a}^{\prime}}\sqrt{-g\left(x_{a}^{\prime}\right)}e^{-2\phi\left(x_{i}^{\prime}\right)}\delta\left(2\phi\left(x_{a}\right)-2\phi\left(x_{a}^{\prime}\right)\right)=V_{d}\sqrt{-g\left(x_{a}\right)}e^{-2\phi\left(x_{a}\right)},\label{eq:reduced non-local dilaton}
\end{equation}

\noindent where we assume the volume of spacetime $V_{d}=\int d^{d}x^{\prime}$
to be finite. Moreover, this volume $V_{d}$
can be absorbed into the definition of $\phi$. Then, the equations
of motion (EOM) of the action (\ref{eq:action with potential}) can
be derived as

\begin{eqnarray}
R_{\mu\nu}-\frac{1}{2}Rg_{\mu\nu}+2\nabla_{\mu}\nabla_{\nu}\phi+\frac{1}{2}g_{\mu\nu}\left[4\left(\partial_{\mu}\phi\right)^{2}-4\nabla^{2}\phi+V\right]-\frac{1}{2}e^{-2\phi}\sqrt{-4\left(\partial_{\mu}\phi\right)^{2}}\gamma_{\mu\nu}I_{1} & = & 0,\nonumber \\
R+4\nabla^{2}\phi-4\left(\partial_{\mu}\phi\right)^{2}-V+\frac{\partial V}{\partial\Phi}-e^{-2\phi}\frac{\gamma_{\mu\nu}\nabla^{\mu}\nabla^{\nu}\phi}{\sqrt{-\left(\partial_{\mu}\phi\right)^{2}}}I_{1}+e^{-2\phi}V^{\prime}I_{2} & = & 0,\label{eq: EoM with potential}
\end{eqnarray}

\noindent with
\begin{eqnarray}
I_{1} & = & -\int d^{D}x^{\prime}\sqrt{-g\left(x^{\prime}\right)}V^{\prime}\left(e^{-\Phi\left(x^{\prime}\right)}\right)\delta\left(2\phi\left(x\right)-2\phi\left(x^{\prime}\right)\right),\nonumber \\
I_{2} & = & \int d^{D}x^{\prime}\sqrt{-g\left(x^{\prime}\right)}\sqrt{-4\partial_{\mu}\phi\left(x^{\prime}\right)\partial^{\mu}\phi\left(x^{\prime}\right)}\delta^{\prime}\left(2\phi\left(x\right)-2\phi\left(x^{\prime}\right)\right),\nonumber \\
\gamma_{\mu\nu} & = & g_{\mu\nu}-\frac{\partial_{\mu}\phi\partial_{\nu}\phi}{\left(\partial_{\mu}\phi\right)^{2}},
\end{eqnarray}

\noindent where $V^{\prime}$ and $\delta^{\prime}$ denote the derivatives
with respect to their arguments ($\frac{\partial}{\partial x}f\left(x\right)\equiv\partial_{x}f\left(x\right)=f^{\prime}\left(x\right)$).
To find the solution of these EOM,
we utilize the ansatz based on the eq. (\ref{eq:inner metric}),
\begin{equation}
{ds^{2}=g_{11}dx^{2}+g_{00}dt^{2}+g_{22}d\varphi^{2}=-dx^{2}+a\left(x\right)^{2}dt^{2}+d\varphi^{2}},\label{eq:ansatz}
\end{equation}
where $0\leq x\leq\frac{\pi}{\lambda}$. By using this
ansatz, the properties of (\ref{eq: EoM with potential}) can be simplified
as

\begin{eqnarray}
e^{-\Phi\left(x\right)} & = &{ V_{d=2}}ae^{-2\phi\left(x\right)},\nonumber \\
I_{1} & = & -\frac{1}{2}{V_{d=2}}a\frac{V^{\prime}}{\phi^{\prime}},\nonumber \\
I_{2} & = & \frac{1}{2}{V_{d=2}}\frac{a^{\prime}}{\phi^{\prime}},\nonumber \\
e^{-2\phi}\sqrt{-4\left(\partial\phi\right)^{2}}I_{1} & = & \frac{\partial V}{\partial\Phi},\nonumber \\
\frac{\gamma_{\mu\nu}\nabla^{\mu}\nabla^{\nu}\phi}{\sqrt{-\left(\partial_{\mu}\phi\right)^{2}}}I_{1} & = & \frac{1}{2}{V_{d=2}}a^{\prime}\frac{V^{\prime}}{\phi^{\prime}}=I_{2}V^{\prime},
\end{eqnarray}

\noindent {where $V_{d}=\int d^{d}x^{\prime}$.} Therefore, the EOM (\ref{eq: EoM with potential}) become

\begin{eqnarray}
R_{00}-\frac{1}{2}Rg_{00}+2\nabla_{0}\nabla_{0}\phi+\frac{1}{2}g_{00}\left[4\left(\partial_{\mu}\phi\right)^{2}-4\nabla^{2}\phi+V\right]-\frac{1}{2}g_{00}\frac{\partial V}{\partial\Phi} & = & 0,\nonumber \\
R_{11}-\frac{1}{2}Rg_{11}+2\nabla_{1}\nabla_{1}\phi+\frac{1}{2}g_{11}\left[4\left(\partial_{\mu}\phi\right)^{2}-4\nabla^{2}\phi+V\right] & = & 0,\nonumber \\
R_{22}-\frac{1}{2}Rg_{22}+2\nabla_{2}\nabla_{2}\phi+\frac{1}{2}g_{22}\left[4\left(\partial_{\mu}\phi\right)^{2}-4\nabla^{2}\phi+V\right]-\frac{1}{2}g_{22}\frac{\partial V}{\partial\Phi} & = & 0,\nonumber \\
R+4\nabla^{2}\phi-4\left(\partial_{\mu}\phi\right)^{2}-V+\frac{\partial V}{\partial\Phi} & = & 0.
\end{eqnarray}

\noindent Substituting the metric (\ref{eq:ansatz}) into the equations
above, we get

\begin{eqnarray}
-2\phi^{\prime2}+2\phi^{\prime\prime}+\frac{1}{2}V-\frac{1}{2}\frac{\partial V}{\partial\Phi} & = & 0,\nonumber \\
2\phi^{\prime2}-2\frac{a^{\prime}}{a}\phi^{\prime}-\frac{1}{2}V & = & 0,\nonumber \\
\frac{2a^{\prime\prime}}{a}-4\frac{a^{\prime}}{a}\phi^{\prime}-4\phi^{\prime\prime}+4\phi^{\prime2}-V+\frac{\partial V}{\partial\Phi} & = & 0.
\end{eqnarray}

\noindent Using (\ref{eq:reduced non-local dilaton}) and (\ref{eq:loop potential}),
we will get the simplified EOM:

\noindent
\begin{eqnarray}
\Phi^{\prime2}-H^{2}+\lambda^{2} & = & \stackrel[k=1]{\infty}{\sum}b_{k}e^{2k\Phi},\nonumber \\
H^{\prime}-H\Phi^{\prime} & = & 0,\nonumber \\
2\Phi^{\prime\prime}-2H^{2} & = & \stackrel[k=1]{\infty}{\sum}2kb_{k}e^{2k\Phi},\label{eq:EoM with potential II}
\end{eqnarray}

\noindent where $H\left(x\right)\equiv\frac{a^{\prime}\left(x\right)}{a\left(x\right)}$.
It is easy to check that when we set all loop corrections $b_{k}=0$,
the EOM (\ref{eq:EoM with potential II}) become the EOM of the action
(\ref{eq:original action}). Moreover, the EOM (\ref{eq:EoM with potential II})
are invariant under the $O\left(d,d\right)$ transformations:

\begin{equation}
a\left(x\right)\rightarrow1/a\left(x\right),\qquad H\left(x\right)\rightarrow-H\left(x\right),\qquad\Phi\left(x\right)\rightarrow\Phi\left(x\right).
\end{equation}

\noindent Now, it is ready to find the exact solution of the EOM (\ref{eq:EoM with potential II}).
After massive attempts, one possible solution of (\ref{eq:EoM with potential II})
is
\begin{eqnarray}
\Phi\left(x\right) & = & \frac{1}{2}\log\left(\frac{1+\gamma}{\sin^{2}\left(\lambda x\right)+\frac{1}{2}\gamma}\right),\nonumber \\
H_{\pm}\left(x\right) & = & \pm\lambda\sqrt{\frac{1+\gamma}{\sin^{2}\left(\lambda x\right)+\frac{1}{2}\gamma}},\label{eq:solution}
\end{eqnarray}

\noindent and

\begin{equation}
b_{1}=0,\qquad b_{2}=-\frac{\frac{1}{2}\gamma\lambda^{2}\left(1+\frac{1}{2}\gamma\right)}{\left(1+\gamma\right)^{2}},\qquad b_{i\geq3}=0,
\end{equation}

\noindent where $H_{\pm}\left(x\right)$ are $O\left(d,d\right)$ dual solutions. In the
perturbative region $\gamma\rightarrow0$, string loop effects disappear
$V\left(e^{-\Phi}\right)\rightarrow-\lambda^{2}$, and the solutions
(\ref{eq:solution}) become,

\begin{eqnarray}
\Phi\left(x\right) & = & -\ln\left(\sin\left(\lambda x\right)\right)+\frac{1}{4}\gamma\left(2-\csc^{2}\left(\lambda x\right)\right)+\mathcal{O}\left(\gamma^{2}\right),\nonumber \\
H_{\pm}\left(x\right) & = & \pm\lambda\csc\left(\lambda x\right)\mp\frac{1}{4}\gamma\lambda\csc\left(\lambda x\right)\left(\csc^{2}\left(\lambda x\right)-2\right)+\mathcal{O}\left(\gamma^{2}\right),
\end{eqnarray}

\noindent which cover the perturbative results (\ref{eq:inner metric}).
For simplicity, we only consider $H_{+}\left(x\right)$, and we denote
$H\left(x\right)$ as $H_{+}\left(x\right)$ in the rest of the paper.
The final result then can be summarized as

\begin{equation}
ds^{2}=-dx^{2}+a\left(x\right)^{2}dt^{2}+d\varphi^{2},\label{eq:regular solution}
\end{equation}

\noindent with

\begin{equation}
a\left(x\right)=c\exp\left(\sqrt{\frac{2\left(\gamma+1\right)}{\gamma}}\mathbb{F}\left(x\lambda,-\frac{2}{\gamma}\right)\right),
\end{equation}

\noindent where $c$ is an integral constant and $\mathbb{F}\left(\phi,m\right)$
is the elliptic integral of the first kind. The regular solution of the metric (\ref{eq:regular solution}) requires the regular Kretschmann scalar: $R^{\mu\nu\rho\sigma}R_{\mu\nu\rho\sigma}=2R^{\mu\nu}R_{\mu\nu}=R^{2}=4\left(H^{2}+H^{\prime}\right)^{2}$.
Therefore, we only need to check whether there is a singularity in the scalar curvature
$R\left(x,\gamma\right)$:

\begin{equation}
R\left(x,\gamma\right)=\frac{2\lambda^{2}\left(2\left(\gamma+1\right)-\sqrt{2}\sin\left(2\lambda x\right)\sqrt{\frac{\gamma+1}{\gamma+2\sin^{2}\left(\lambda x\right)}}\right)}{\gamma+2\sin^{2}\left(\lambda x\right)}.
\end{equation}

\begin{figure}[H]
\begin{centering}
\includegraphics[scale=0.6]{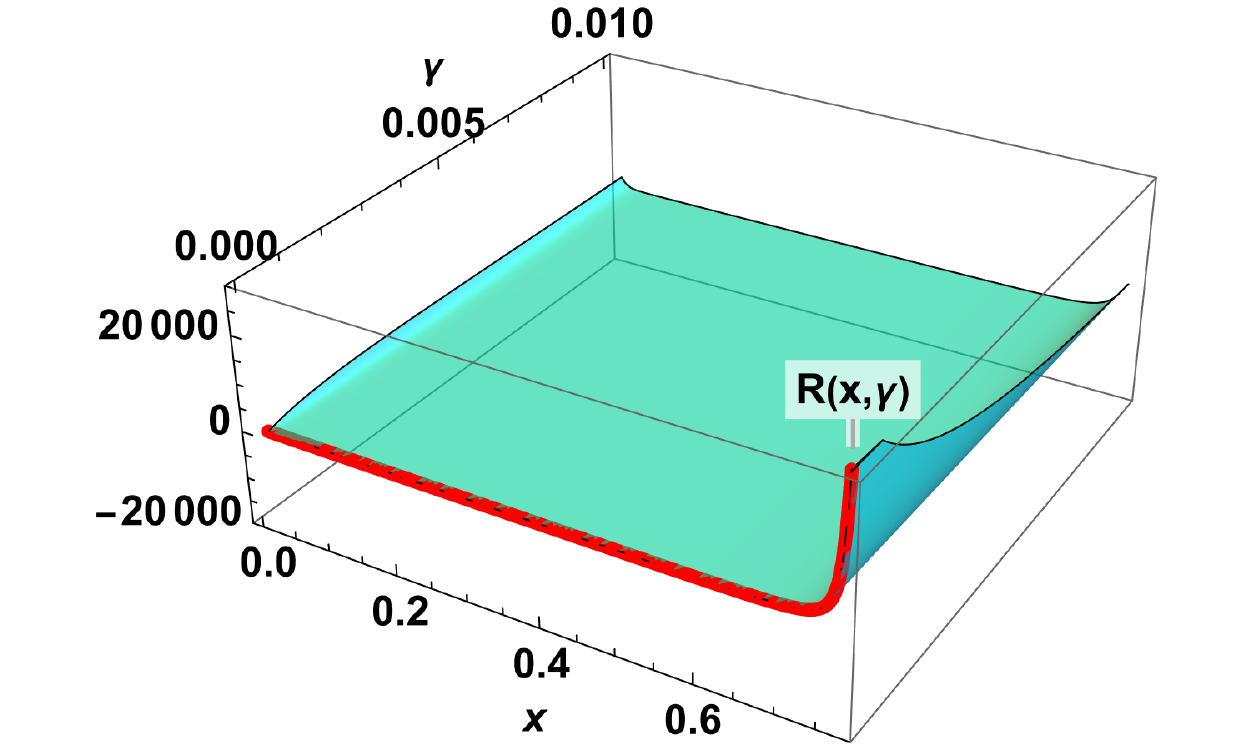}
\par\end{centering}
\centering{}\caption{\label{fig:scalar curvature all region} The behavior of scalar curvature
$R\left(x,\gamma\right)$, where $\lambda=4$ and $0\protect\leq x\protect\leq\frac{\pi}{\lambda}\simeq0.8$.
The solid red line describes the scalar curvature $R_{0}\left(x\right)$
of the perturbative solution (\ref{eq:inner metric}).}
\end{figure}

\noindent It is easy to see that this result is regular everywhere in the region $0\leq x\leq\frac{\pi}{\lambda}$
when $\gamma$ is non-zero. The behavior of this result can be plotted
in figure (\ref{fig:scalar curvature all region}). When $\gamma$ increases, string loop effects matter, and
then the spacetime singularity disappears.

\noindent

\noindent On the other hand, we consider the behavior of the physical dilaton $\phi$, which can be obtained from the equation (\ref{eq:odd invariant dilaton}):

\begin{eqnarray}
\phi\left(x\right) & = & \frac{1}{2}\left(\Phi+\ln\sqrt{\left|g_{00}\times g_{22}\right|}\right),\nonumber \\
 & = &\sqrt{\frac{\left(\gamma+1\right)}{2\gamma}}\mathbb{F}\left(x\lambda,-\frac{2}{\gamma}\right)+\frac{1}{2}\log c+\frac{1}{4}\log\left(\frac{\gamma+1}{\frac{\gamma}{2}+\sin^{2}\left(\lambda x\right)}\right).
\end{eqnarray}

\noindent This solution is regular in the region $0\leq x\leq\pi/\lambda$, which can be seen in the picture as follow:

\noindent
\begin{figure}[H]
\begin{centering}
\includegraphics[scale=0.6]{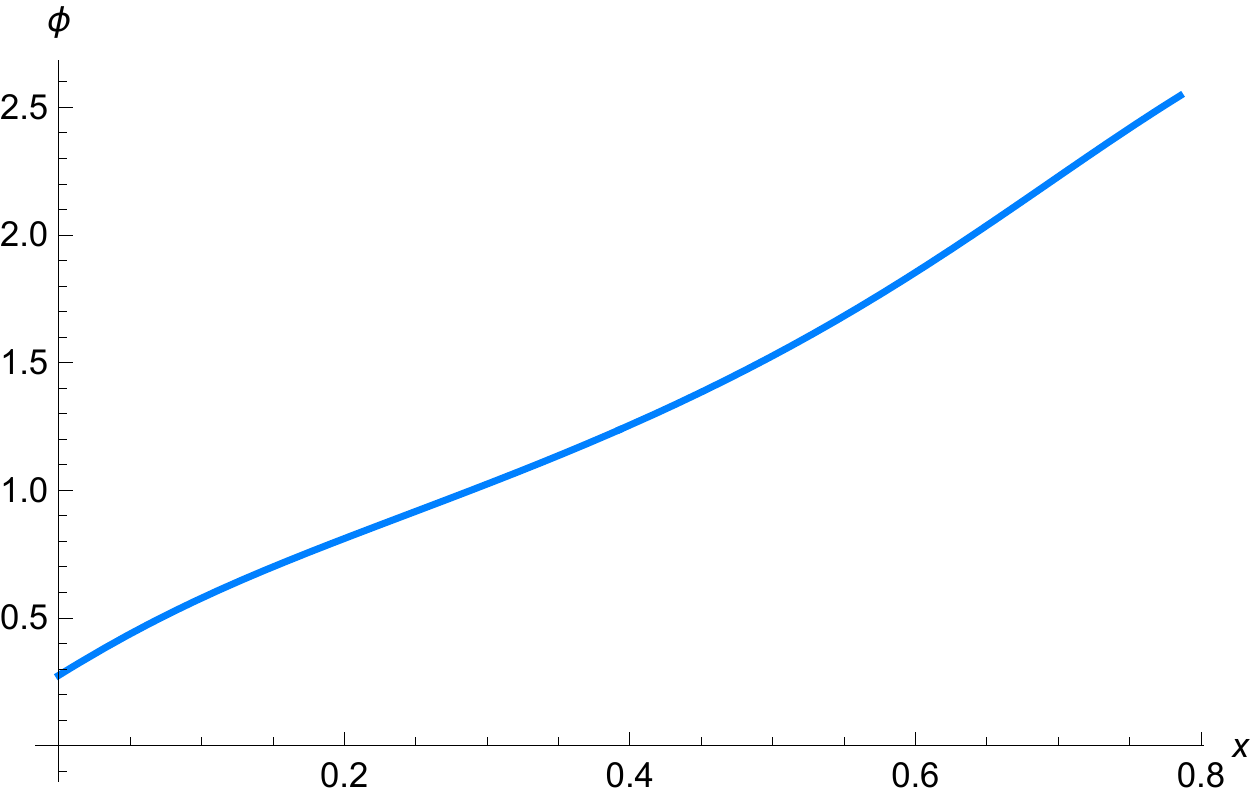}
\par\end{centering}
\centering{}\caption {\label{fig:physical dilaton} The behavior of the physical dilaton $\phi\left(x\right)$, we choose the parameters $\gamma=2$, $\lambda=4$ and $c=1$. }
\end{figure}

\section{Discussion}

Based on the most general potentials for the string loop corrections (\ref{eq:loop potential}),
we can obtain some new realizations of black string. Let
us recall the singular black string metric (\ref{eq:inner metric}):

\begin{eqnarray}
ds_{II}^{2} & = & -dx^{2}+\tan^{2}\left(\frac{\lambda}{2}x\right)dt^{2}+d\varphi^{2},\nonumber \\
\Phi & = & -\ln\left(\sin\left(\lambda x\right)\right).
\end{eqnarray}

\begin{figure}[H]
\begin{centering}
\includegraphics[scale=0.6]{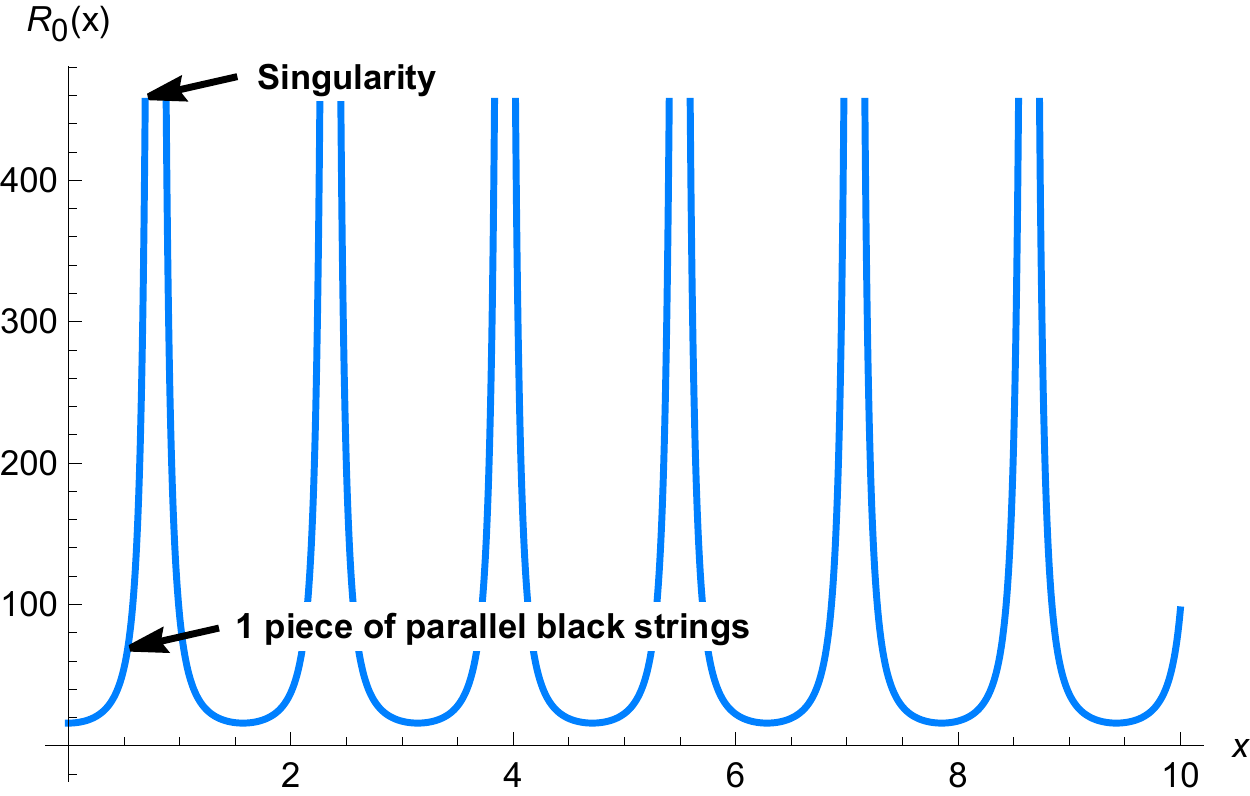}
\par\end{centering}
\centering{}\caption{\label{fig:para strings} We only pick one of black string solutions, the singularities
locate at $n\frac{\pi}{\lambda}$.}
\end{figure}

\noindent We have mentioned before, this metric is periodic with $\frac{n\pi}{\lambda}$
where $n=1,3,5,\ldots$ is an odd number. In the traditional case,
only $0\leq x\leq\frac{\pi}{\lambda}$ makes sense, because the spacetime
singularity locates at $\frac{\pi}{\lambda}$. To see it clear, let
us recall the coordinate transformation

\begin{equation}
\frac{r}{M}=\cos^{2}\left(\frac{\lambda}{2}x\right),
\end{equation}

\noindent where the region $0\leq r\leq M$ corresponds to $\left(n-1\right)\frac{\pi}{\lambda}\leq x\leq n\frac{\pi}{\lambda}$.
It implies that we only select one interval of all $n=1,3,5,\ldots$ solutions, see figure (\ref{fig:para strings}).

\noindent

\begin{figure}[H]
\begin{centering}
\includegraphics[scale=0.6]{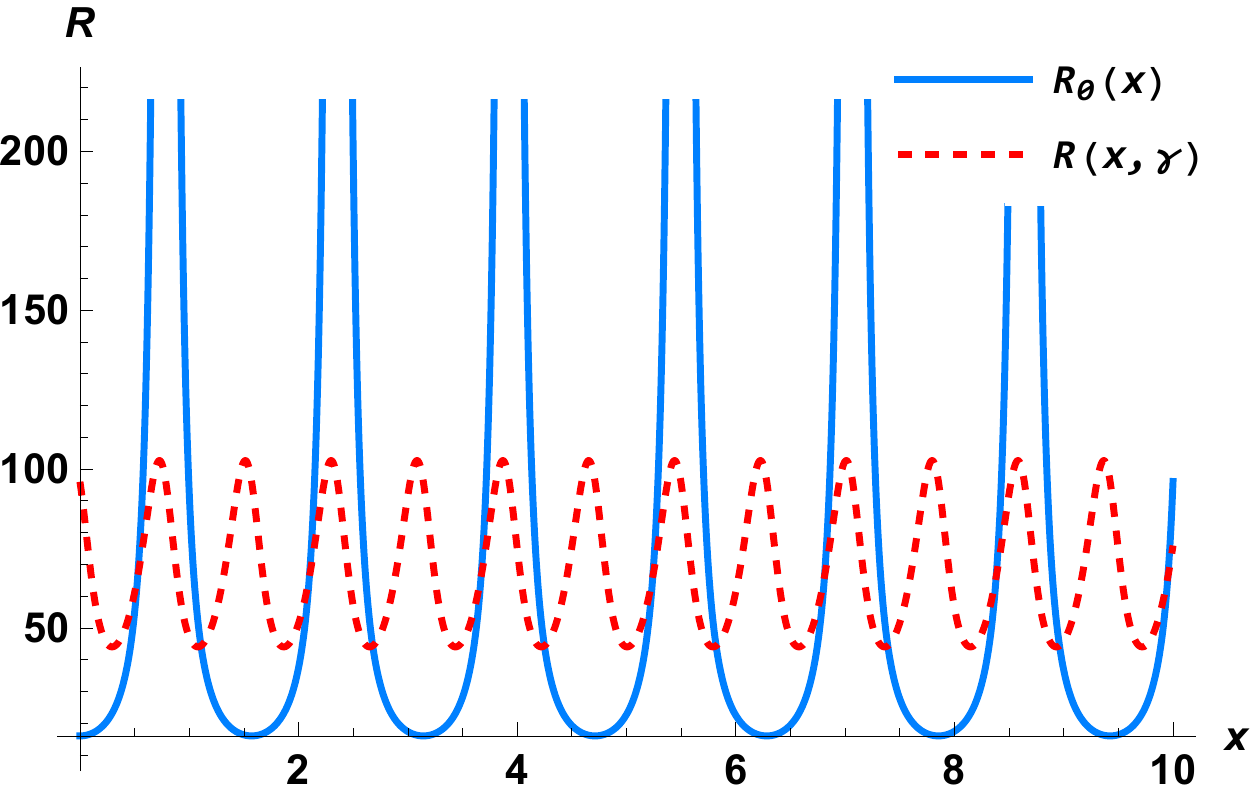}
\par\end{centering}
\centering{}\caption{\label{fig:para strings 2} The blue solid line denotes the
scalar curvature of the perturbative solution and the dashed red line
represents the non-perturbative regular solution.}
\end{figure}

\noindent However, when we introduce the loop corrections, it removes
all singularities in figure (\ref{fig:para strings}). Then the originally unrelated solutions smoothly connect with each other.
Therefore, it introduces new spacetime structure and extends the interval from
$0\leq x\leq\frac{\pi}{\lambda}$ to $x\geq0$, see figure
(\ref{fig:para strings 2}).

\noindent

In short, in this paper, we constructed the most general potentials for the string
loop corrections. Based on this potential, we solved the equations
of motion of three dimensional low energy effective action exactly.
The exact solution showed that the black string singularity could
be resolved by the loop corrections. In the perturbative region, this
exact solution reduced to the three dimensional singular black string
solution.  In the following work, it is worthwhile to study how to obtain other exact solution (which possesses two spacetime singularities) of coset model \cite{Sfetsos:1992yi} from the loop corrected action.

\vspace{5mm}

{\bf Acknowledgements}
We thank the useful discussions with Xin Li, Peng Wang, Houwen Wu and Haitang Yang. This work is supported in part by the NSFC (Grant No. 12105031).

\end{document}